# Revisiting Fizeau's Observations: Spectral study of Na source using Newton's rings

K S Umesh[#], Denny Melkay, J Adithya, Sai Prem Shaji, N Ganesh,
R Dharmaraj, Rajkumar Jain, S Nanjundan

\# Author for correspondence: ksumesh@sssihl.edu.in

Physics Department,
Sri Sathya Sai Institute of Higher Learning,
Brindavan Campus, PO Kadugodi,
Bangalore, 560067

## Abstract

Conventional Newton's rings is generally used in undergraduate laboratory to determine the average wavelength of Na doublet. A minor modification of this setup enables us to view simultaneously Newton's rings both in reflected as well as transmitted light. A movable glass plate with respect to the plano-convex lens allows us to observe the variation of contrast/visibility of these fringes and thus allows us to determine the separation of the yellow doublet.

# 1. Introduction:

Newton's rings experiment is a classic example of fringes of equal thickness or Fizeau fringes[1,2]. This is generally used in undergraduate optics laboratory, as it enhances conceptual understanding of internal and external reflections and associated phase shifts. It also requires fair degree of experimental skill to perform this. Even in the literature[3,4,5,6], we find it being used for classroom demonstrations, for determination of wavelength of He-Ne laser etc. In this work we demonstrate how this could also be used to estimate the doublet separation apart from being used to determine the average wavelength.

# 2. Theory of formation of Newton's rings:

For a film of thickness, t, the path difference between the two partially reflected light from top and the bottom surface of film is given by[7]:

$$\delta = 2\, n_f\, t\, \cos(\theta_t)$$

where $n_f$ :Refractive index of film. t:Thickness of glass plate, $\theta_t$:Angle of refraction

If the film is of varying thickness t, the optical path difference varies even without variation in the angle of incidence. Thus if the direction of the incident beam is fixed, say at normal incidence, a dark or bright fringe will be associated with a particular thickness for which λ satisfies the condition for destructive or constructive interference, respectively. For this reason, fringes produced by variable thickness film are called Fringes of equal thickness or Fizeau fringes. When an air wedge formed between the spherical surface and an optical flat is illuminated from a laser or sodium vapour lamp, equal thickness contours for a perfectly spherical surface are formed. These circular fringes are called Newton's rings. These fringes are formed around the point of contact. At the centre, t=0, the path difference between reflected rays is λ/2 as a result of reflection. The centre of the fringe pattern (zero[th] order fringe) thus appears dark.

By finding the radius of two or more rings, we can get the wavelength of light using the formula,

$$\lambda = \frac{r_{m+p}^2 - r_m^2}{pR}$$

where, $r_m$ & $r_{m+p}$ radii of the $m^{th}$ ring and $(m+p)^{th}$ ring

R, Radius of curvature of the lens

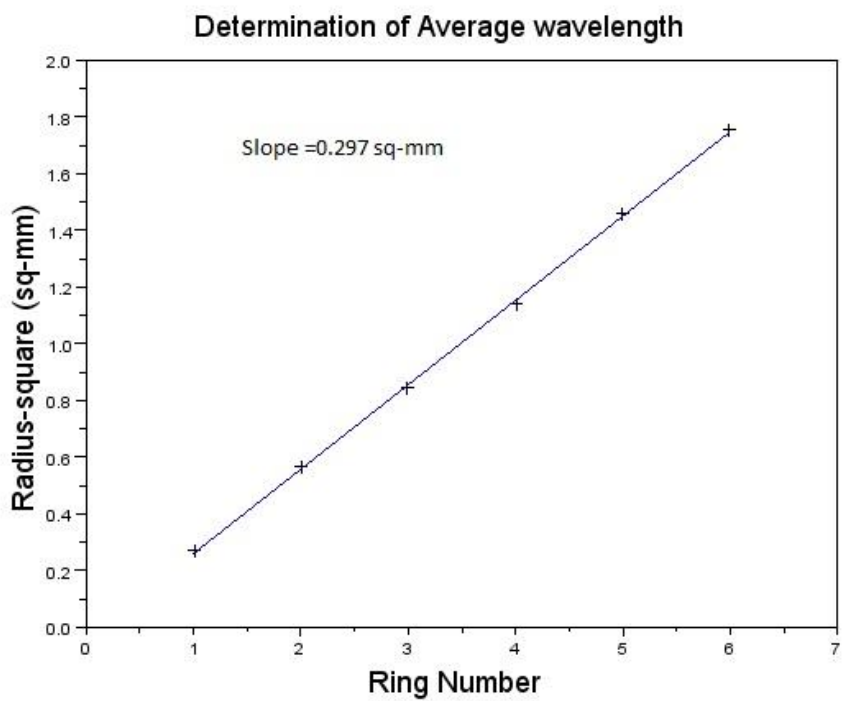

Fig(1) : Graph of ring no versus diameter-square to determine the average wavelength

The conventional Newton's rings experiment was performed and the graph of ring no. m versus $r_{m+p}^2$ was plotted. The slope of this graph is $(r_{m+p}^2 - r_m^2)/p$

Therefore the average wavelength of sodium source

= slope /radius of curvature (R)

= 0.297 x10$^{-6}$ m$^2$ / 0.5 m

= 595.3 nm

## 3. Finding the Fine structure using Newton's rings

We know that the sodium lamp emits two wavelengths: 5890A and 5896A. These two wavelengths form two sets of Newton's rings with coincident centres. If we examine a few rings near the point of contact of lens and glass plate, the two sets of rings appear to coincide; but if they are traced to a sufficient distance from the centre, the misalignment becomes more and more apparent.

Consequently, after some path difference, the bright fringe of one set of rings will occupy the same position as the dark fringe of the other set, and they will mutually annhilate to a uniform intensity. Continuing the same line of reasoning, it is evident that perfect coincidence and perfect misalignment of the two systems of rings would recur alternately at regular intervals[8].

To observe this variation of contrast, the glass plate is made movable in a perpendicular direction. This would result in linear variation of path difference. As the higher order fringes appear, due to finite temporal coherence, the contrast variation can be clearly seen.

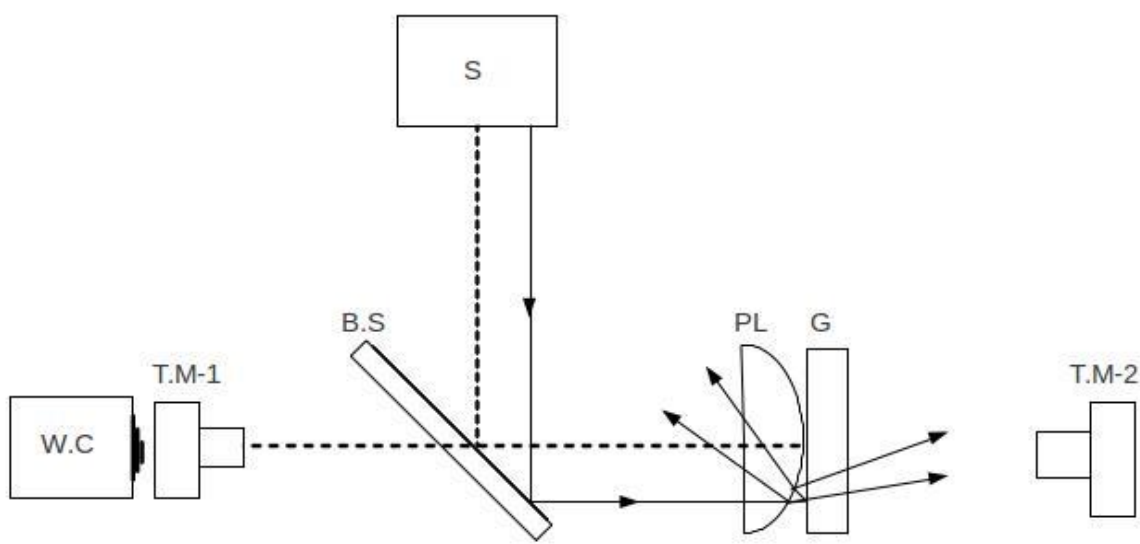

Fig(2) : Schematic diagram of Newton's rings with glass plate, G and plano-convex lens, PL mounted vertically and the Glass plate made movable. T.M-1 and T.M-2: Travelling microscopes, W.C: Webcam connected to travelling microscope

## 4. Analysis:

The separation of two wavelengths present in the source, used to observe interference is related to the path difference, as in the case of Michelson interferometer[7]. If Δλ is the separation between two spectral lines and λ is the average wavelength, then

$$\Delta\lambda = \frac{\lambda^2}{2\,\Delta d}$$

where Δd mirror movement required between two consecutive coincidences in a Michelson interferometer. This formula can be used in Newton's rings in the case where glass plate is moved with respect to the plano-convex lens. In this case, Δd represents distance through which the glass plate is moved from zero path difference(glass plate touching the plano-convex lens) position to the subsequent region of maximum contrast.

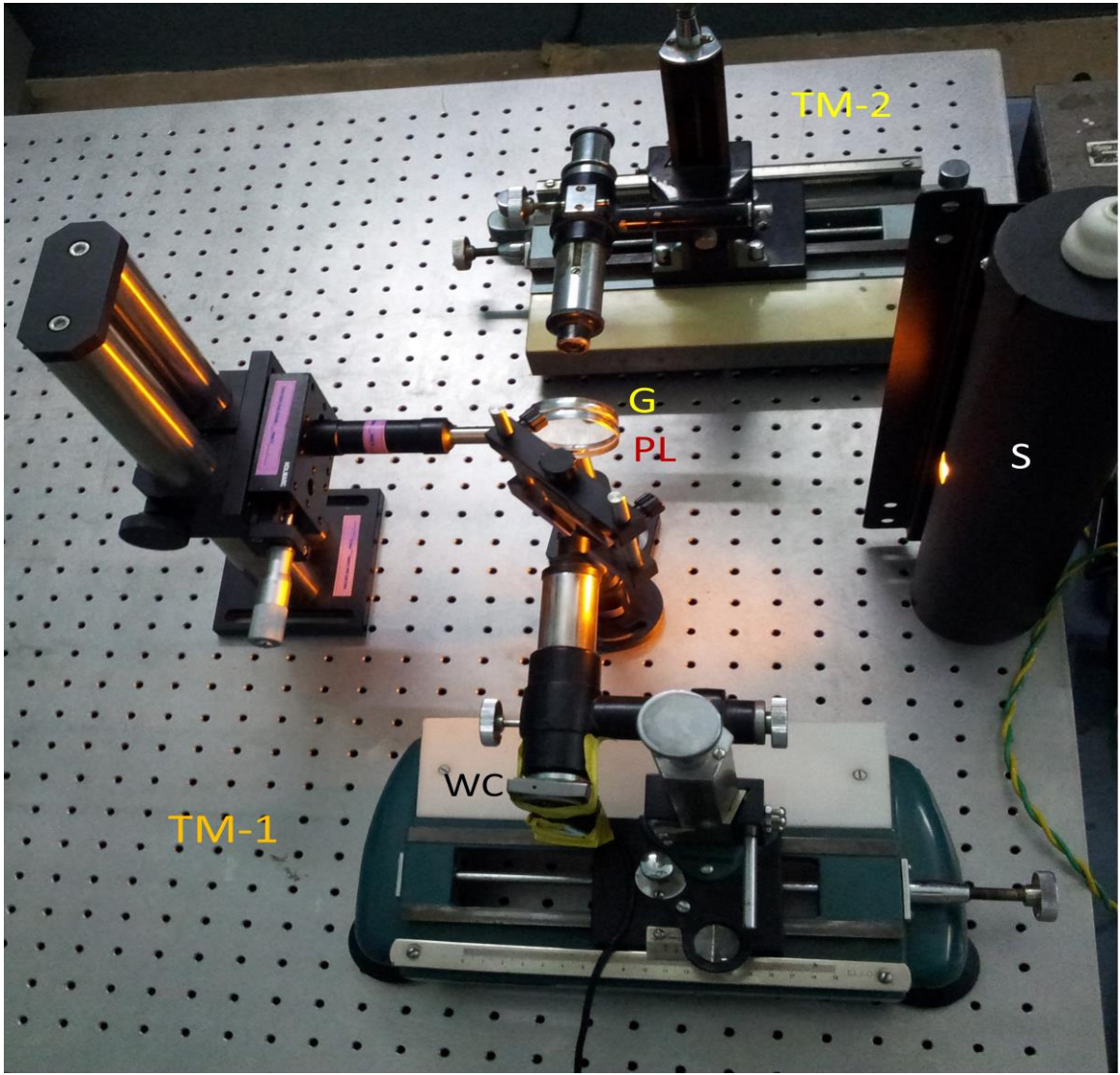

Fig(3): Photograph of experimental setup

## 5. Experimental details and Results:

The schematic diagram and the photograph of the experimental setup are shown in fig(2) and fig(3) respectively. The glass plate was mounted on a micrometer translation stage so that its distance can be varied in a controlled manner. Newton's rings in the reflected light were captured using a webcam (without the lens) attached to travelling microscope without the eyepiece. Newton's rings were captured for every 10µm distance. From the snapshot, the visibility of the fringes was calculated using ImageJ software. Sample snapshots are at different distances are provided in fig(5) to fig(9) which clearly show the contrast variation. Fig(10) & fig(11) show the Newton's rings in the reflected and transmitted light.

Plot of distance versus visibility is shown in the fig(4). As expected, there is a periodic variation of contrast starting with its maximum value at zero path difference. From the graph, we get Δd = 250μm.

$$\Delta\lambda = \frac{\lambda^2}{2\,\Delta d}$$

$$= (595.3\text{nm})^2/2(250\mu m)$$

$$= 7\,\text{A}$$

This shows that the separation of the Na doublet whose expected value is 6A, could be estimated to an accuracy of about 17 percent using this analysis.

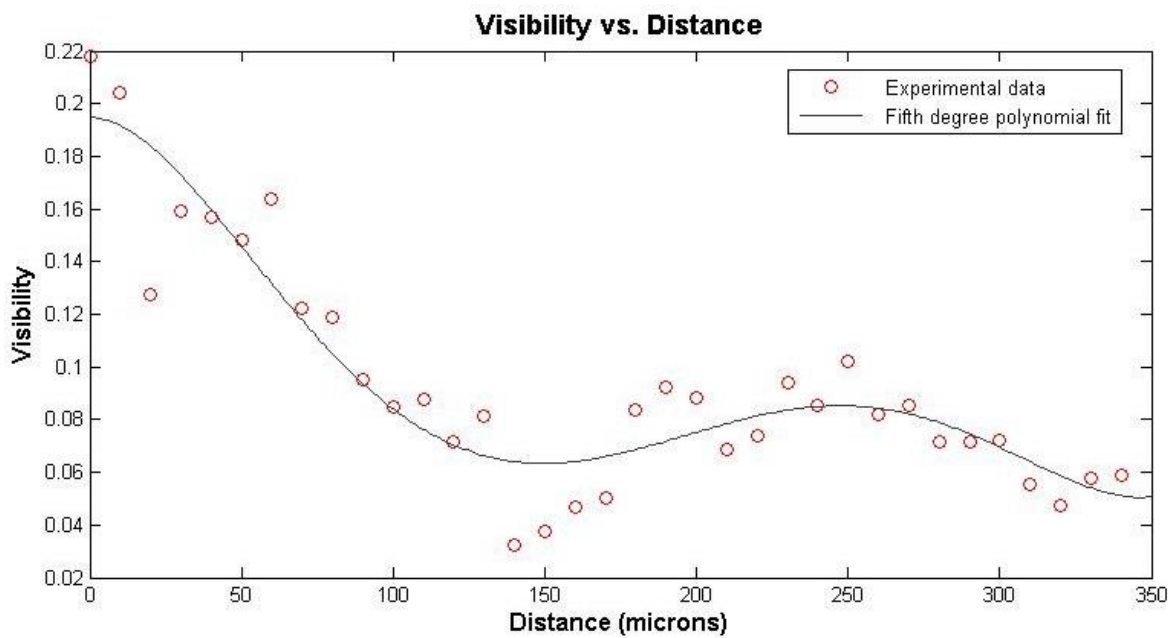

Fig(4) : Plot of visibility as a function of distance of glass plate from the plano-convex lens.

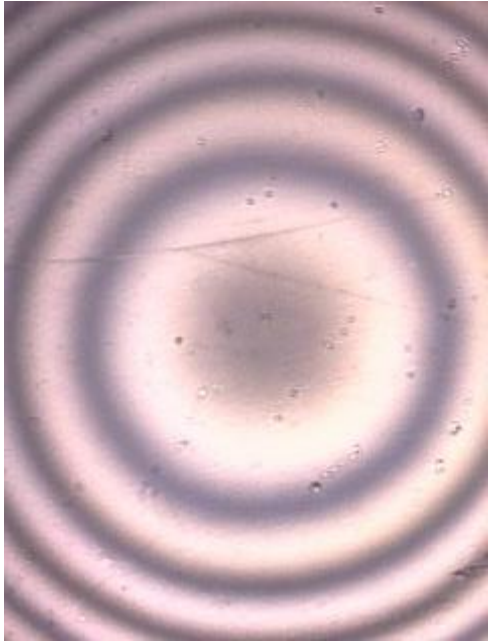

**Fig. 5: Zero path difference**

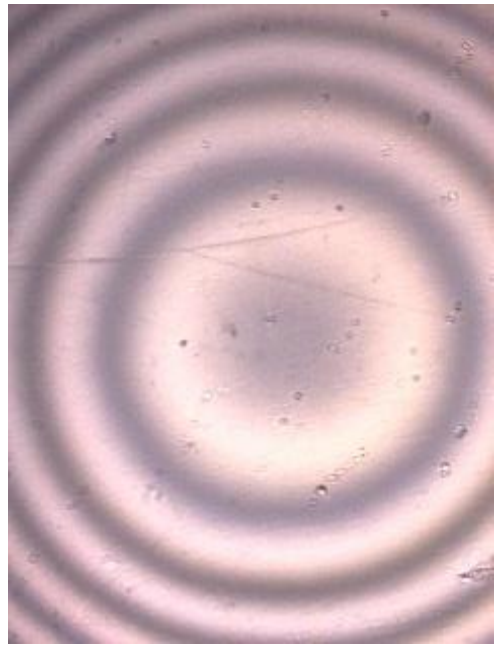

**Fig. 6: Path difference = 50micron**

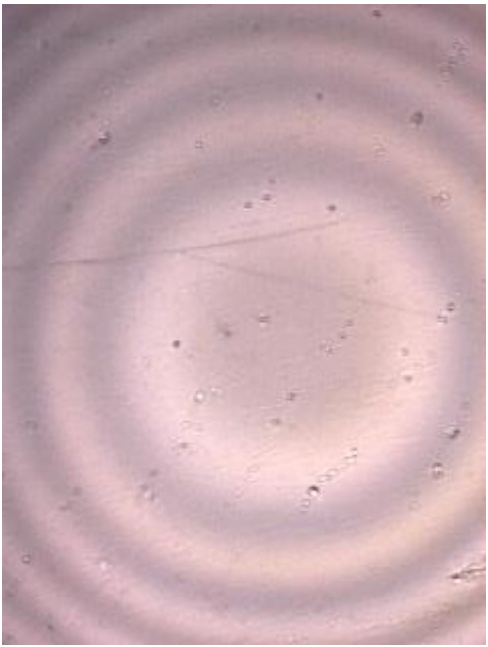

**Fig. 7: Path difference=100 micron**

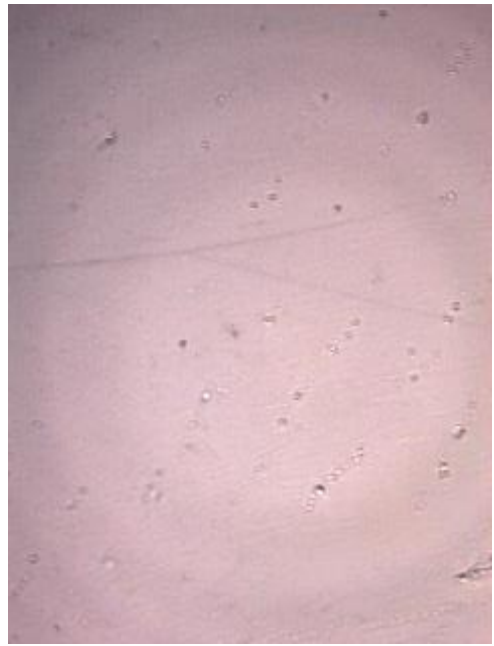

**Fig. 8: Path difference=150 micron**

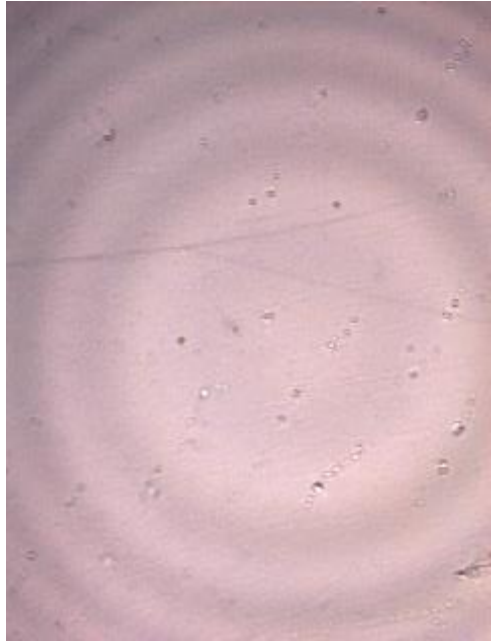

**Fig. 9: Path difference=250 micron**

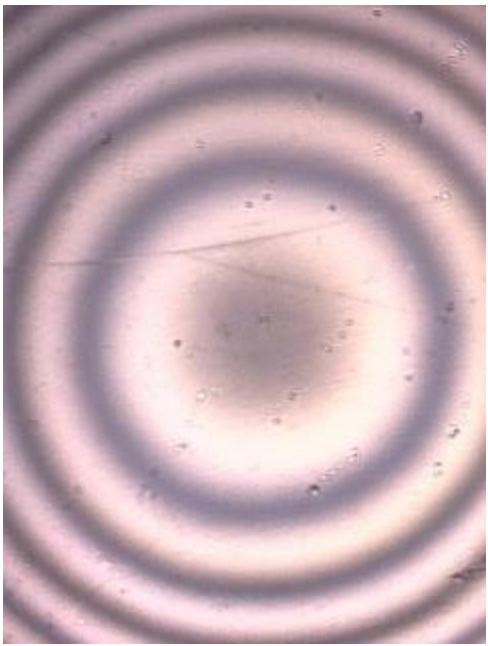

**Fig. 10: Newton's rings: Reflected light**

- **Central Dark Fringe**
- **High Contrast**

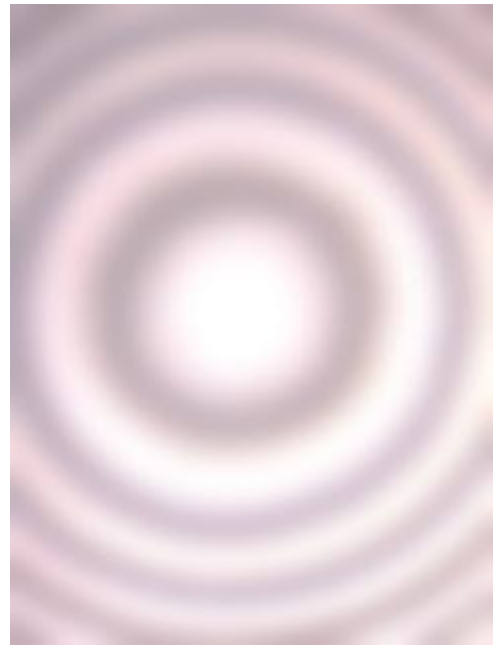

**Fig. 11: Newton's rings: Transmitted light**

- **Central Bright Fringe**
- **Low Contrast**

## 6. Conclusion:

In this work, we have revisited Fizeau's observations by modifying the conventional Newton's rings setup. Both the glass plate and plano-convex lens are mounted vertically which enables us to view both the reflected and transmitted rings simultaneously, compare and contrast them with considerable ease. Moreover, the glass plate mounted on a micrometer translation stage, allows us to observe clearly the variation fringe contrast/visibility. From the graph of distance versus visibility, we were able to calculate the separation of the Na doublet.

Acknowledgements:

Authors express their thanks to VGST, DST GoK for awarding the CISE 2011-12 project, which enabled us to carry out this work in the Physics Department, SSSIHL, Brindavan Campus